\documentstyle[]{mn}
\newif\ifAMStwofonts
\def\ion#1#2{#1\,{\sevensize\uppercase\expandafter{\romannumeral#2}}} 
\def\markcite#1{}
\def\micron{\hbox{$\mu$m}}
\def\mum{\hbox{\micron~}}

\def\lsun{\hbox{L$_\odot$}}

\ifoldfss
  \ifCUPmtlplainloaded \else
    \NewTextAlphabet{textbfit} {cmbxti10} {}
    \NewTextAlphabet{textbfss} {cmssbx10} {}
    \NewMathAlphabet{mathbfit} {cmbxti10} {} 
    \NewMathAlphabet{mathbfss} {cmssbx10} {} 
  \fi
  \ifAMStwofonts
    \ifCUPmtlplainloaded \else
      \NewSymbolFont{upmath} {eurm10}
      \NewSymbolFont{AMSa} {msam10}
      \NewMathSymbol{\upi}     {0}{upmath}{19}
      \NewMathSymbol{\umu}     {0}{upmath}{16}
      \NewMathSymbol{\upartial}{0}{upmath}{40}
      \NewMathSymbol{\leqslant}{3}{AMSa}{36}
      \NewMathSymbol{\geqslant}{3}{AMSa}{3E}

      \let\leq=\leqslant 
       
    \fi
  \fi
\fi 

\ifnfssone
  \newmathalphabet{\mathit}
  \addtoversion{normal}{\mathit}{cmr}{m}{it}
  \addtoversion{bold}{\mathit}{cmr}{bx}{it}
  \newmathalphabet{\mathbfit} 
  \addtoversion{normal}{\mathbfit}{cmr}{bx}{it}
  \addtoversion{bold}{\mathbfit}{cmr}{bx}{it}
  \newmathalphabet{\mathbfss} 
  \addtoversion{normal}{\mathbfss}{cmss}{bx}{n}
  \addtoversion{bold}{\mathbfss}{cmss}{bx}{n}
  \ifAMStwofonts
    \ifCUPmtlplainloaded \else
      \UseAMStwoboldmath
      \makeatletter
      \new@mathgroup\upmath@group
      \define@mathgroup\mv@normal\upmath@group{eur}{m}{n}
      \define@mathgroup\mv@bold\upmath@group{eur}{b}{n}
      \edef\UPM{\hexnumber\upmath@group}
      \new@mathgroup\amsa@group
      \define@mathgroup\mv@normal\amsa@group{msa}{m}{n}
      \define@mathgroup\mv@bold\amsa@group{msa}{m}{n}
      \edef\AMSa{\hexnumber\amsa@group}
      \makeatother
      \mathchardef\upi="0\UPM19
      \mathchardef\umu="0\UPM16
      \mathchardef\upartial="0\UPM40
      \mathchardef\leqslant="3\AMSa36
      \mathchardef\geqslant="3\AMSa3E

      \let\leq=\leqslant 

    \fi
  \fi
\fi 

\ifnfsstwo
  \DeclareMathAlphabet{\mathbfit}{OT1}{cmr}{bx}{it}
  \SetMathAlphabet\mathbfit{bold}{OT1}{cmr}{bx}{it}
  \DeclareMathAlphabet{\mathbfss}{OT1}{cmss}{bx}{n}
  \SetMathAlphabet\mathbfss{bold}{OT1}{cmss}{bx}{n}
  \ifAMStwofonts
    \ifCUPmtlplainloaded \else
      \DeclareSymbolFont{UPM}{U}{eur}{m}{n}
      \SetSymbolFont{UPM}{bold}{U}{eur}{b}{n}
      \DeclareSymbolFont{AMSa}{U}{msa}{m}{n}
      \DeclareMathSymbol{\upi}{0}{UPM}{"19}
      \DeclareMathSymbol{\umu}{0}{UPM}{"16}
      \DeclareMathSymbol{\upartial}{0}{UPM}{"40}
      \DeclareMathSymbol{\leqslant}{3}{AMSa}{"36}
      \DeclareMathSymbol{\geqslant}{3}{AMSa}{"3E}

      \let\leq=\leqslant 

    \fi
  \fi
\fi 

\ifCUPmtlplainloaded \else
  \ifAMStwofonts \else 
    \def\upi{\pi}
    \def\umu{\mu}
    \def\upartial{\partial}
  \fi
\fi

\title[NGC 253]{8-13 \mum spectroscopy of NGC 253:\\  a spatially resolved starburst}
\author[C. C. Dudley \& C. G. Wynn-Williams]
       {C. C. Dudley \& C. G. Wynn-Williams\\
        Institute for Astronomy, University of Hawaii, 2680 Woodlawn Dr., Honolulu, Hawaii 96822, USA }
\date{}
\pubyear{1998}

\begin{document}

\maketitle


\begin{abstract}

NGC 253 is a nearby spiral galaxy that is currently
undergoing a nuclear burst of star formation in a 100 pc
diameter region.  We present spatially resolved 8--13 \mum
low-resolution spectra at four positions along the ridge of
8--13 \mum emission.  We find that the relative strengths of
the ionic, dust emission features, and dust continuum emission
vary with position in the galaxy but can be accounted for
everywhere without recourse to extinction by silicates.  The
brightest mid-infrared peak (which is displaced from the
nucleus) has elevated levels of both continuum and 11.1--12.9
\mum ``plateau'' emission, indicative of dust heated within a
photo-dissociation region.  Spectra obtained over the course
of 3 yr at the position of the brightest mid-infrared peak
show no significant time variation.

\end{abstract}


\section{Introduction\label{ch3:int}}

NGC 253 is a nearby ($D$ $\sim$3 Mpc; Tully\markcite{n253-tul88}
1988) S(B)bc infrared bright galaxy whose infrared luminosity
($\sim$2 $\times 10^{10}$ \lsun; Telesco \&
Harper\markcite{n253-te80} 1980) is powered by a $\sim$100 pc
diameter burst of star formation in the circumnuclear region.
The infrared luminosity of NGC 253 exceeds that of the Milky
Way by a factor of 2 with a larger relative fraction arising
from the nuclear region. It thus provides an important
proving ground to test our understanding of luminous infrared
galaxies that are powered by compact starbursts.  In these
sources the higher luminosity per unit volume and the higher
average density in the ISM may produce effects that are infrequently
observed in star-forming regions in the Milky Way.

The central regions of NGC 253 are strongly affected by dust
extinction at visual wavelengths.  It is now generally agreed
that the nucleus of the galaxy lies near to the point marked TH2
(Turner\markcite{n253-tu85} \& Ho 1985) in Fig. 1, some 3\arcsec~
(45 pc) away from the brightest infrared peak P1 (Sams et
al.\markcite{n253-sa94} 1994; Kalas \&
Wynn-Williams\markcite{n253-ka94} 1994; Pi\~na et
al.\markcite{n253-pi92} 1992; Keto et al.\markcite{n253-ke93}
1993).  The nature of P1 is unclear.  Based on its mid-infrared
properties, Kalas \& Wynn-Williams\markcite{n253-ka94} (1994)
suggested that it is a very young supernova remnant embedded in
dust, while Watson et al. \markcite{ngc253-wa96}(1996), who
observed it with the {\sl Hubble Space Telescope}, suggested that
it contains a young star cluster.

The 8--13 \mum spectral region provides useful information on the
conditions in starbursts through thermal dust continuum emission,
dust emission features, and ionic emission
lines.  The five strongest of the dust emission features occur at 3.3,
6.2, 7.7, 8.6, and 11.3 \micron.  The features are frequently
attributed to polycyclic aromatic hydrocarbons (PAHs) (Puget \&
L\'eger\markcite{n253-pu89} 1989), but other
laboratory analogues have also been proposed  
(e.g., Duley\markcite{n235-du89} 1989; Sakata \& Wada\markcite{n253-sak89}
1989; Ellis et al.\markcite{n253-el94} 1994).  PAH emission is widespread in the Galaxy.
The features are observed in photo-dissociation regions (PDRs), reflection nebulae,
and infrared cirrus, as well as in planetary nebulae. The ionic emission lines [\ion{Ar}{3}], [\ion{S}{4}], and
[\ion{Ne}{2}], on the other hand, all arise in
\ion{H}{2} regions and can be used as indicators of both the
hardness and strength of the ionizing continuum radiation in
obscured regions.

Roche \& Aitken\markcite{n253-14} (1985) used a 7\farcs6 aperture
to obtain 8--13 \mum spectra at two positions in the starburst
in the nucleus of NGC 253.  They found spectra that are broadly
similar to those of many other star-forming galaxies, with
contributions from dust continuum, PAH emission, and
[\ion{Ne}{2}].  They discuss evidence for silicate absorption in
their spectra, a topic that we re-examine in this paper.

In this paper we present new 8--13 \mum spectra at four positions
in the starburst nucleus along the mid-IR ridge of emission
with particular emphasis on the peak P1, for which we have
obtained 
observations over a 3-years baseline.

\section{Observations and Data Reduction\label{ch3:odr}}

\subsection{Observations\label{ch3:o}}

The data were obtained using the cooled grating 10 and 20 \mum
spectrometer CGS3 (Cohen
\& Davies\markcite{n253-coh95} 1995) at the United Kingdom Infrared Telescope
(UKIRT) on Mauna Kea.    We used low-resolution
mode (${\lambda\over\Delta\lambda}\sim$60) so that the entire
8--13 \mum atmospheric window was observed.  Two separate
interleaved grating positions are
required to fully sample a spectrum with the 32-element
linear detector array.  Observations
were performed by chopping with a 15\arcsec~ throw in the
northwest-southeast direction at a frequency of 5 Hz for 10
s, followed by a beam switch of equal duration. No more than 10
such cycles were performed before observing at the second
grating position.

\subsection{Pointing and astrometry\label{ch3:pa}}

Fig. 1 shows the positions of our spectroscopic observations
labeled as P0, P1, P2a, and P2b.  These observations were
obtained on three nights, one each in 1992, 1993, and 1994.
Table 1 is a log of the observations giving the date, aperture
size in arcseconds for full width at 10 per cent power, per-detector element
on-source integration times in seconds, and for P0, P2a, and
P2b the relative offset from the P1 observation in
arcseconds.  In Fig. 1, only the 2\farcs4 aperture is
shown for P1. The central position of the CGS3 aperture with
respect to the optical guide camera was established by using
an automatic peak up routine on bright stars.  In 1992, 1993,
and 1994 the stars were HR 188, HR 8892, and HR 268,
respectively, which are all within 10$^\circ$ of NGC 253.  The
telescope coordinate system was established by observing 2 to
3 Carlsburg Meridian Circle stars near the galaxy.  In 1992,
and 1994, we then slewed to P1, while in 1993, we peaked up
manually on the P1 using the integrated flux between 8 and 13
\micron. We estimate that the 1992 and 1994 slew positions are
accurate to 1\farcs5, while the offsets from the position of
P1 are accurate to 0\farcs5.

The publication of the {\sl Hipparcos} catalog has allowed a
more precise determination of the position of the star used
as an astrometric reference by Kalas \&
Wynn-Williams\markcite{n253-ka94} (1994).  In Table 2 we give
the proper-motion-corrected position for SAO 166579 based
on the {\sl Hipparcos} catalog.  The position has not been
corrected for parallax, which is the main source of the quoted
error.  The conversion to
the FK4 system was performed using the routine BPRECESS
provided with IDL version 5 using the epoch keyword.  The
position is consistent with the proper motion corrected
position derived from the FK4 format PPM catalog
(available from the NASA Astrophysics Data System), which
is 0\fs02 east and 1\farcs34 south of the SAO position
used by Kalas \& Wynn-Williams\markcite{n253-ka94}
(1994).  Table 2 also gives the revised positions for
Kalas \& Wynn-Williams's Peaks 1--3, the radio positions of
TH2 and TH7 (Turner \& Ho\markcite{n253-tu85} 1985), and the positions of two red objects
observed by Watson et al.\markcite{n253-wa96} (1996).  In
Table 2 we see that the revised position for Kalas \&
Wynn-Williams's Peak 1 is now consistent with TH7 and that
their Peak 2 is now about 1 $\sigma$ distant from
TH2.  The positions of Kalas \& Wynn-Williams' Peaks 1 and
3 are consistent with Watson et al.'s Bright blob and Spot
$a$, respectively; if Peak 1 and their Bright blob are brought into
coincidence, then the positions of Peak 3 and Spot $a$
agree within 0\fs01 in $\alpha$ and 0\farcs02 in $\delta$,
within the estimated internal accuracy of the two
observations.

\subsection{Data Reduction\label{ch3:dr}}

Our spectra have been flux calibrated against bright stars.  In
1992, 1993, and 1994, we used HR 1457 (N=$-$3.03;
Tokunaga\markcite{n253-to84} 1984), HR 7001 (N$\equiv$0.00), and
HR 7949 (N=0.0; assumed), respectively.  Wavelength-dependent
variations in instrumental response and atmospheric transmission
were removed by dividing the object spectra by stellar spectra
and multiplying by a model Planck function.  In 1992 and 1993,
all spectra were first corrected to 1 airmass using atmospheric
transmission curves based on data collected during each observing
run.  In 1994, the same star was used as a spectral and flux
calibrator and was observed within 0.1 airmass of the source. In
1992, HR 0617 was used for spectroscopic calibration with an
assumed temperature of 4780 K; in 1993, HR 0188 was used with an
assumed temperature of 4690 K; and for the 1994 observations, the
assumed temperature for HR 7949 was 4690 K. All the spectral
standards are spectral type K0 or earlier so that they should not
be strongly affected by SiO absorption (Cohen et
al.\markcite{n253-coh92a} 1992a).  Deviation from our model
Planck function due to wavelength-dependent H$^-$ opacity is also
likely to be small (Cohen et al.\markcite{n253-coh92b} 1992b).
We estimate our absolute photometric accuracy to be better than 10
per cent.

Wavelength calibration was based upon observations of a Kr
lamp as described by Hanner, Brooke \&
Tokunaga\markcite{n253-ha95} (1995).

Due to year-to-year variations in instrumental setup, which
resulted in slightly different wavelength scales for our
spectra, we have resampled all our spectra to a common
wavelength scale using Gaussian weighting with $\sigma=0.06$
\micron.  Thus, the individual data presented below are nearly
but not entirely independent of their spectral neighbors.  We
found no significant difference between the spectra obtained
at P1 over the 3 yr, so we have combined them to enhance
the signal-to-noise ratio.  Given the compact size $\leq
1\farcs2$ (Keto et al.\markcite{n253-ke93} 1993) of P1 between 8
and 13 \micron, it is not surprising that spectra obtained
with our two aperture sizes should agree, and this agreement
can be taken as evidence that P1 was reasonably well centered
all 3 yr.  

\section{Results\label{ch3:r}}

In Fig. 2 we present our spectra of the starburst region
of NGC 253.  The panels are labeled in accordance with
Fig. 1 and Table 1.  We note that certain features are
apparent in all the spectra.  The brightest emission
features are the [\ion{Ne}{2}] forbidden line at 12.8 \mum
and the PAH emission feature at 11.3 \micron.  The
8.6-\mum PAH feature is also present in each spectrum.  The
continuum emission has a broad minimum near 10 \micron; we
shall argue in Section 4.1 that this `continuum' is actually a
combination of a true thermal continuum between 10 and 13
\mum and the long wavelength shoulder of the 7.7-\mum PAH
emission feature.  Each spectrum also shows signs of an
elevated level of emission between 11.1 and 12.9 \micron.
This so-called ``plateau'' emission has been previously
observed in the Orion Bar photo-dissociation region (PDR) by
Roche, Aitken \& Smith \markcite{n253-ro89}(1989).  There
are no signs of the 10.5-\mum [\ion{S}{4}] or 9.0-\mum
[\ion{Ar}{3}] forbidden emission lines in any spectrum.

As a check on our calibration, we have computed the average
flux density between 11.15 and 12.25 \mum of the bright
source P1.   Our value of 4.7$\pm$0.1 Jy is in good agreement
with the estimate of Pi\~na et al.\markcite{n253-pi92} (1992)
of 4.8 Jy.

\section{Discussion\label{ch3:d}}

Our discussion will first concentrate on two progressively more
detailed fits to our spectra culminating with a table of
measurements of spectral feature properties.  We will then
discuss the question of silicate absorption in NGC 253,
followed by an analysis of the strengths of the ionic emission
lines.  Finally we discuss the possible origins of the
continuum emission and the spatial variations seen in our
spectra. 

We fit the features in our spectra in two stages.  First we apply
a two-component model to the whole of each spectrum consisting of
a theremal continuum underlying PAH emission.  This procedure
allows us to test whether or not silicate absorption is present
in these spectra by examining the agreement of the observations
with what the fits predict near 8 \micron.  Second, we decompose
the emission above the thermal continuum longward of 11 \mum
into three components: PAH feature emission, [\ion{Ne}{2}]
emission, and plateau emission.

\subsection{Initial fitting: Continuum and 11.3 \mum feature\label{ch3:d1}}

Our initial procedure is to fit the 10--13 \mum spectrum of each
region to a combination of thermal blackbody and 11.3-\mum PAH
emission feature, and then compare this model's predicted 8--10
\mum emission with the observations.  The procedure is similar to
that employed for Arp 299C in our earlier paper (Dudley \&
Wynn-Williams\markcite{n253-dud93} 1993) except that the
continuum is now estimated solely from the spectra themselves
rather than in conjunction with longer wavelength photometric
data.  For a PAH template we use the Orion Bar data of Roche et
al. \markcite{n253-ro89}(1989) degraded in spectral resolution to
match the present data.  The particular spectrum we use is that
taken at Beckin et al.'s \markcite{n253-be76}(1976) position 4,
following Roche \markcite{n253-ro89a}(1989) so that a prediction
can be made of the strength of 8--10 \mum emission.

In Fig. 2, the solid lines are Planck function fits to the
continuum between 10 and 11 \mum and at 13.2 \mum and the
dotted line shows the Orion Bar spectrum scaled so that the
intensity of Orion Bar emission around the 11.3-\mum PAH
feature between 10.9 and 11.7 \mum matches that found above
the continuum fits in our spectra.  The shaded regions in
Fig. 2 represent the reasonable range in the predicted
8--10 \mum emission based on the 10--13 \mum fits.  They are
bounded on the bottom by the Orion Bar spectrum of Bregman et
al.\markcite{n253-br89} (1989) with the same scaling relative
to the Roche et al.\markcite{n253-ro89} (1989) spectrum as
that adopted by Roche\markcite{n253-ro89a} (1989), and on the
top by the same multiplied by a factor of 2.  The shaded
region reflects the observation that there is a factor of 2
scatter in the ratio of the intensities of the 11.3 to
7.7+8.6 \mum features reported by Zavagno, Cox, \&
Baluteau\markcite{n253-za92} (1992) in observations of
galactic \ion{H}{2} regions.  Of the sources they analyze,
the Orion Bar is among those with the strongest 11.3-\mum features
relative to the 7.7+8.6-\mum features.  As seen in Fig. 2,
the simple model described in the last paragraph provides an
acceptable description of the gross features of all the
spectra of NGC 253 we obtained.  We therefore conclude,
leaving aside the ionic [\ion{Ne}{2}] emission line, that the
8--13 \mum spectrum NGC 253 can be explained by a thermal
continuum rising towards longer wavelengths, plus a PAH
spectrum scaled to the intensity of the 11.3-\mum feature.
Thus the minimum near 9.7
\mum need not be produced by silicate absorption; we discuss
the limits on silicate absorption in Section 4.3.

The color temperatures and 13.2-\mum continuum flux densities of
the four measured regions are given in rows 1--2 of Table 3.
As a check on the appropriateness of modeling the continuum in
this manner, we extrapolate the results of the fits to 19.5
\mum and compare the extrapolations with the data given by
Pi\~na et al. (1992).  For IRS 1, they report a 19.5 \mum flux
density of 24.8 Jy, which is within 6\% of our extrapolation
at P1 of Fig. 1.  For all the other positions in Fig. 1, our
extrapolations over predict the contours of their 19.5 \mum
deconvolved map.  The largest over predictions (a factor of 2)
occur at P0 and P2b where we measure the lowest values of the
color temperature.  There are several possible explanations
for this.  First, notwithstanding Pi\~na et al.'s efforts to
deconvolve their map, there may remain important differences
between their beam size and ours.  Second, a blackbody may not
be the best representation for the continuum; a $\lambda^{-1}$
or $\lambda^{-1.5}$ emissivity law gives a much better fit to
their map.  Third, some silicate absorption may have led to an
underestimate of the continuum temperature at these positions.
The silicate optical depth required ($\tau_{\rm sil}(9.7) \sim$
0.7, with $T_{\rm cont}=150$ K) does not exceed the limits
we set on the silicate optical depth in section 4.2.

\subsection{Detailed fitting: The plateau emission}

In the Orion Bar, the [\ion{Ne}{2}]
emission, 11.3-\mum PAH feature, and plateau all have different
spatial distributions (Roche et
al.\markcite{n253-ro89} 1989); we have chosen to fit these three
components separately after first subtracting the thermal
continuum. 

Our fitting method is quite similar to that employed by Roche et
al.\markcite{n253-ro89} (1989) to analyse the Orion Bar.  There
are two main differences: (1) In the Orion Bar, a feature at 12.7
\mum shows the same spatial variation as the 11.3-\mum feature
(Roche et al.\markcite{n253-ro89} 1989), but due to our lower
spectral resolution, the 12.7-\mum feature cannot be
independently measured here.  We have therefore assumed a fixed
ratio between the intensities of the 11.3- and 12.7-\mum features
of 2.4 based on the Orion Bar results.  (2) We have chosen a
plateau spectral profile that is stronger at longer wavelengths
in $F_\lambda$ rather than the flatter profile employed by Roche
et al.\markcite{n253-ro89} (1989).  While our profile has the
same spectral width as theirs, it rises linearly in wavelength
and is twice as strong at 12.5 \mum than at 11.5 \micron.  This
shape was chosen to match the shape of the plateau emission in
the P1 spectrum, which has the highest signal-to-noise ratio.

We first fitted and subtracted the plateau by matching
intensity between 11.6 and 12.2 \mum for the data and the
profile.  We then matched intensities and subtracted an
assumed Gaussian profile centered at 11.3 \mum and with
$\sigma$ = 0.13 \micron, where the matching occurred between
11.0 and 11.4 \micron; a Gaussian centered on 12.7 \mum with
width $\sigma$ = 0.26 \mum was also subtracted as described
previously.  Finally the intensity of the [\ion{Ne}{2}] feature was
measured between 12.6 and 13.0 \micron.  These procedures left
residuals that were not significantly different from zero
except in the P1 and the P2a spectra, where an apparent feature
at 11.5 \mum of intensities 1.3 $\pm$ 0.4 and 1.4 $\pm$ 0.4
$\times 10^{-15}$ W m$^{-2}$, respectively, measured between 11.3
and 11.7 \mum remained.  Since such a feature has been
observed in some Galactic sources exhibiting PAH emission
(Roche, Aitken \& Smith\markcite{n253-ro91a} 1991a), these
features could be real, but since we have modified the shape
of our plateau model from that employed by Roche et
al.\markcite{n253-ro89} (1989), we do not make any strong claim
for this.  This fitting procedure also produced a lower
estimate of the 11.3-\mum intensity than the model used to
produce Fig. 2.  The largest reduction occurred for P1, where
the new 11.3-\mum intensity is about 80 per cent of that used in
Fig. 2.  About half of this reduction is due to 
plateau subtraction, and half is due to our effort to exclude
flux from the apparent feature at 11.5 \micron.

We summarize the results of the fits shown in Fig. 3 in
rows 3--5 of Table 3.  Rows 6 and 7 are limits on the
intensities of two additional ionic emission lines: 8.99-\mum
[\ion{Ar}{3}], and 10.5-\mum [\ion{S}{4}].  The equivalent
widths of the 11.3-\mum PAH feature and the 12.8-\mum
[\ion{Ne}{2}] line in rows 8 and 9 are based on the
intensities in rows 3 and 5 and a model continuum that
includes contributions from both the thermal continuum given
in rows 1 and 2 and a plateau contribution, and in the case
of [\ion{Ne}{2}], an assumed 12.7-\mum feature contribution.
Row 10 gives the intensity ratios between the plateau and
11.3-\mum PAH features.  The
errors given in Table 3 are 1 $\sigma$ while upper limits are
3 $\sigma$.

There are few spectra of galaxies available with sufficient
signal-to-noise ratios to detect the 11.1--12.9 \mum plateau
feature, so it is difficult to make comparisons with other
galaxies, although plateau emission is clearly present in
the two {\sl Infrared Space Observatory} spectra of the
antennae galaxy system presented by Vigroux et
al. \markcite{n253-vi96} (1996).

Most of what is known about the plateau emission is based on
observations of the Orion Bar region.  Roche et
al.\markcite{n253-ro89} (1989) have made spatially resolved
observations of the Orion Bar.  Their observations of the ratio
of the intensities of the plateau emission to 11.3-\mum feature
varies between 1.5 and 6.7.  The lowest value is found on the
visible Bar itself (Becklin et al.'s (1976) position 4), which
also behaves to some degree as a reflection nebula, and the
highest value is found at a position 20\arcsec~ south, well
within the dense part of the molecular cloud.  At the position
where the plateau emission is the strongest (5\arcsec~ south of
position 4), the ratio has a value of 2.7.  The range of values
for this ratio listed in Table 3 for NGC 253 falls within the
range of values we have just calculated.  Of particular interest
are the high values of the ratio seen at P1 and P2a that can
only be approximated by the southern most values in the Orion
Bar.  These values may show that conditions
similar to those occurring in the deep layer of the Orion Bar PDR
are more prevalent than those occurring adjacent to the ionization
front.

\subsection{Limits on Silicate Absorption}

Aitken \& Roche\markcite{n253-ai84} (1984) have proposed a
method of determining the amount of silicate absorption
required for spectra that include PAH emission based on
observations of galactic planetary nebulae and employing the
equivalent width of either the 11.3- or 8.6-\mum features.
Given the factor of 2 dispersion in the observed 11.3- to
7.7+8.6-\mum features intensity ratio in galactic \ion{H}{2} regions,
we prefer to set an upper limit on the amount of true thermal
continuum that might be present at 8
\mum and thus an upper limit on $\tau_{\rm Sil}$(9.7) where
$$\tau_{\rm
Sil}(9.7)=\ln\left({F_\lambda(8)+F_\lambda(13)\over 2\times
F_\lambda(9.7)}\right)$$ (Aitken \&
Jones\markcite{n253-ai73} 1973). We therefore take the
difference between the lower boundary of the shaded region at 8
\mum (less our model continuum at 8 \micron) and the spectra
as a rough limit on the amount of continuum that can
contribute to the flux at 8 \micron.  This value, along with
the model continuum flux at 9.7 and 13 \micron, combine to give
an upper limit on $\tau_{\rm Sil}$(9.7) in the equation of
Aitken \& Jones\markcite{n253-ai75} (1973).  These limits
are given in row 11 of Table 3.

We find that our upper limits on $\tau_{\rm Sil}$(9.7) are in
agreement with the $\tau_{\rm Sil}$(9.7) reported by Roche \&
Aitken\markcite{n253-ro85} (1985) of 1.4 $\pm$ 0.3 based on the
method of Aitken \& Roche\markcite{n253-ai84} (1984) for NGC
253, given the very different aperture sizes we have
employed.  We note that these limits on $\tau_{\rm Sil}$(9.7)
apply to silicate absorption by cold dust, but not to the
optical depth in dust that is emitting between 8--13 \micron.
The limit for P1 is consistent with the amount of extinction
estimated by Watson et al.\markcite{n253-wa96} (1996) based on
$V-I$ colors under the assumption that $A_V=15\times\tau_{\rm
Sil}$(9.7) (Aitken \markcite{n252-ai81}1981).

\subsection{Ionic Lines}

The presence and absence of ionic lines in our spectra allow
us to derive properties of the ionized gas within NGC 253.

In all four of the spectra shown in Fig. 2, the [\ion{Ne}{2}]
emission line is clearly detected, but the [\ion{Ar}{3}]
and [\ion{S}{4}] lines are not.  Of these two lines the limits on
I([\ion{S}{4}]) in Table 3 are probably more reliable given the
systematic errors introduced by both the presence of the
shoulder of the 7.7+8.6-\mum features and the adjacent telluric
O$_3$ absorption that could affect the continuum estimate for
the 8.99-\mum [\ion{Ar}{3}].  We therefore concentrate on 
what can be learned from the limits on I([\ion{S}{4}]) emission.
The ratio of I([\ion{S}{4}]) to I([\ion{Ne}{2}]) is sensitive to the
hardness of the Ly$_{\rm cont}$ radiation field ionizing the
\ion{H}{2} regions producing the emission and thus to the
presence of very hot stars.  Assuming a number abundance ratio
of S to Ne of 0.27, n$_e$ $\sim 10^3$--$10^4$ cm$^{-3}$, and
Ly$_{\rm cont}$ fluxes following the models of
Kurucz\markcite{n253-ku79} (1979), Rubin\markcite{n253-ru85}
(1985) gives ratios of I([\ion{S}{4}])/I([\ion{Ne}{2}]) of 0.02
and 1 for exciting stars of $T_{\rm eff}$ = 33,000 and 36,000 K,
respectively.  Our upper limits on I([\ion{S}{4}]) imply that
I([\ion{S}{4}])/I([\ion{Ne}{2}]) $<$ 0.33, 0.09, 0.14, and 0.13
for each of the spectra P0, P1, P2a, and P2b, respectively.  Thus
the nondetection of the [\ion{S}{4}] line suggests that the bulk of the
stars producing ionizing radiation have $T_{\rm eff}$ $<$ 36,000
K.  The strongest limit is seen at P1, where a finer grid of
models than those examined here would be useful.  The limit on
$T_{\rm eff}$ holds even if $\tau_{\rm Sil} \sim$ 1.2, which
would increase our upper limits on
I([\ion{S}{4}])/I([\ion{Ne}{2}]) by a factor of 1.7 based on the
Mathis\markcite{n253-32} (1990) extinction curve with $R\sim3$
supplemented by a $\mu$ Cep silicate profile matching our
spectral resolution (see Dudley \&
Wynn-Williams\markcite{n253-dud97} 1997 for details.)

The [\ion{Ne}{2}] emission may be compared with
observations at other wavelengths.  Puxley \&
Brand\markcite{n253-pux95} (1995) measure a Br$\gamma$ flux of
$0.18\times10^{-15}$ W m$^{-2}$ at P1 using a 3\arcsec~ slit
oriented along the galaxy's major axis.  This gives
I(Br$\gamma$)/I([\ion{Ne}{2}]) = 0.015 which is within 25 per cent of
the prediction of Dudley \& Wynn-Williams\markcite{n253-dud93}
(1993) of 0.011 but is quite low compared to the models of
Rubin\markcite{n253-ru85} (1985) in the range of $T_{\rm eff}$
suggested by the upper limit on I([\ion{S}{4}]), which
predict a ratio of 0.048.  Different assumptions about Ne
abundance, Brackett to Balmer series intensity ratios, and the
[\ion{Ne}{2}] collision strength can account for about half
of the discrepancy between the two predictions.  On the other
hand, applying the reddening deduced by Watson et
al.\markcite{n253-wa96} (1996) towards P1 to the prediction of
Rubin\markcite{n253-ru85} (1985) gives good agreement with the
observed ratio.  Applying the I([\ion{Ne}{2}]) to
I(Br$\gamma$) conversion factor of Dudley \&
Wynn-Williams\markcite{n253-dud93} (1993) followed by a
further conversion using the Br$\gamma$ intensity to 6-cm
free-free flux density ratio of 8 $\times10^{11}$ Hz from
Wynn-Williams\markcite{n253-wy82} (1982) to the observed
[\ion{Ne}{2}] intensity predicts a 6-cm flux density of 17
mJy, which is not too different from the observed value of
about 10 mJy for TH 7 (Ulvestad \& Antonucci\markcite{n253-ul97} 1997).  The 6 to
2 cm spectral index of this source is $\alpha=-0.24\,\,
(S_\nu\propto\nu^\alpha)$, intermediate between optically
thin free-free and synchrotron indices, suggesting that both
processes contribute to the radio emission.

For P2a, at the nucleus of the galaxy, the predicted Br$\gamma$
flux based on the [\ion{Ne}{2}] flux (Dudley \&
Wynn-Williams\markcite{n253-dud93} 1993) is
$0.10\times10^{-15}$ W m$^{-2}$, which is a factor of 2 lower
than the $0.24\times10^{-15}$ W m$^{-2}$  observed by Puxley \& Brand\markcite{n253-pux95}
(1995).  This may indicate that there is an ionizing source
associated with the nonthermal radio source (TH2) that
produces Br$\gamma$ emission that is not mirrored in the
[\ion{Ne}{2}] emission, possibly because the ionized gas density
is too high (n$_e$ $>$ $10^5$ cm$^{-3}$) to produce as much
[\ion{Ne}{2}] emission.  This speculation is not supported,
however, when using the ratio of Rubin\markcite{n253-ru85}
(1985) which overpredicts the observed Br$\gamma$ flux by a
factor of 2.  As in the case of P1, this effect might also be
accounted for by reddening.  Both comparisons are beset by
possible aperture differences, an unknown Ne abundance, and
uncertainties in determining the [\ion{Ne}{2}] intensity due to
our subtraction of an assumed 12.7-\mum PAH feature so that only
firm conclusion that can be drawn is that the Br$\gamma$ and
[\ion{Ne}{2}] are in agreement within a factor of 4 under
the assumption that they both arise from \ion{H}{2} regions.

The 6-cm radio continuum in P2a is about 40 mJy (Ulvestad \&
Antonucci\markcite{n253-ul97} 1997), stronger than the free-free emission
predicted from the intensities of either [\ion{Ne}{2}] or
Br$\gamma$, which amount to about 13 mJy.  This is consistent
with the statement of Turner \& Ho (1985) that the 2-cm radio
continuum emission from TH2 must be due to nonthermal emission
since its brightness temperature of 10$^5$ K is too high to be
due to free-free emission from \ion{H}{2} regions.   On the other
hand, the amount of free-free emission that we would predict by
substituting the [\ion{Ne}{2}] to Br$\gamma$ conversion factor
given by Rubin (1985) but retaining the Br$\gamma$ to 6-cm flux
density factor adopted from Wynn-Williams (1982) gives good
agreement with the observed 6-cm emission.

\subsection{Continuum emission}

In standard Case B recombination theory, radio free-free
emission, Brackett series recombination line emission, and
[\ion{Ne}{2}] emission all trace the number of ionizing
photons with little sensitivity to electron temperature and
density for the values typical of \ion{H}{2} regions.  Thus,
[\ion{Ne}{2}] emission may be used to predict the amount of
grain emission due to trapped Ly$\alpha$ heating in
\ion{H}{2} regions while avoiding the possible problems of
extinction at short wavelengths and nonthermal emission at
radio wavelengths.  Taking the Br$\gamma$-to-[\ion{Ne}{2}] intensity
ratio predicted by Dudley \&
Wynn-Williams\markcite{n253-dud93} (1993) and a Br$\gamma$
intensity to 6-cm free-free flux density ratio from
Wynn-Williams\markcite{n253-wy82} (1982), we predict a
free-free flux density at 6 cm of between 5 and 17 mJy for
the four spectra.  Using the continuum temperatures given in
Table 3, we may then use formula A4 of Genzel et
al.\markcite{n253-ge82} (1982) to predict the expected flux
density at 13.2 \mum due to trapped Ly$\alpha$ heating of
dust and compare these predictions with the observed
values given in Table 3.  We find that the observed continuum
exceeds these predictions by large factors of 30--50.  We
therefore conclude that Ly$\alpha$ heating fails to account
for the 10--13 \mum continuum in any of our spectra.

\subsection{Spatial Variations\label{ch3:d2}}

Before discussing the possible reasons for spatial variations
among our spectra of NGC 253 we need to review the status of
the compact source P1.

In this paper we have already shown that the 8--13 \mum spectrum
of P1 is consistent with a combination of thermal dust emission
and Orion Bar-like emission along with [\ion{Ne}{2}] emission
similar to that seen in starburst galaxies (see Fig. 2).
Further, we observed no time variation in the 8--13 \mum spectrum
over the 3 yr that we observed it.  In view of the absence of
time variations, and of Watson et al.'s (1996) {\sl HST}
detection of a star cluster spatially coincident with P1, the
suggestion of Kalas \& Wynn-Williams (1994) that P1 is a
dust-embedded recent supernove is probably incorrect.

Our spectrum of P1 can provide a new estimate of the 1--1000 \mum
luminosity of the source based on the color temperature and flux
of the continuum. A 145-K blackbody with a 13.2-\mum flux density
of 6.7 Jy at a distance of 3 Mpc has a luminosity of 1.6$\times
10^9$ L$_\odot$.  We recall the agreement noted in section 4.1 between this extrapolation and the 19.5 \mum data of Pi\~na et al. (1992).
This
luminosity estimate for P1 is in good agreement with the estimate of
Watson et al.\markcite{n253-wa96} (1996) of 1.8$\times 10^9$
$L_\odot$ which is based on their estimate of the fraction of
the flux due to P1 in the 12.5-\mum map of Keto et
al.\markcite{n253-ke93} (1993) applied to the $L_{10-350 \mum}$
for the inner 30\arcsec~ of NGC 253 given by Telesco \&
Harper\markcite{n253-te80} (1980).  However, our estimate of
the flux due to P1 in the Keto et al.\markcite{n253-ke93}
(1993) map is a factor of 2.3 larger than that of Watson et
al.\markcite{n253-wa96} (1996), so the agreement is largely
fortuitous.  By either estimate, P1 provided $\sim$8 per cent of the
infrared luminosity of the galaxy.  

The 11.3-\mum feature is clearly present in all our spectra but
shows a large variation in EW.  This is consistent with the
observations of the related 3.3-\mum feature reported by Kalas \&
Wynn-Williams\markcite{n253-ka94} (1994), where it was found to
be relatively weak compared to the continuum at P1.  The EW of
the 11.3-\mum feature ranges over a factor of 4 and is
anti-correlated with the continuum temperature, while the ratio
of the intensities of [\ion{Ne}{2}] to 11.3-\mum feature cover a
factor of 2 and all fall within the range reported by Roche et
al.\markcite{n253-ro91b} (1991b) for 19 galaxies showing both
features.

The young star cluster model for P1 leads, at least
qualitatively, to a plausible explanation for the differences
seen between the spectum at P1 (and to some extent P2a) and
elsewhere in the galaxy.  Because the interstellar radiation
field is presumably stronger within the compact P1 star
cluster than the interstellar radiation field elsewhere,
regular dust grains will be heated to higher temperatures as
is observed (Table 3).  Since we have already shown that
Ly$\alpha$ heating is insufficient to heat the dust, we need
to invoke direct dust heating by starlight.  The prominence
of the plateau emission in the spectrum of P1 suggests that
emission that arises in PDRs is particularly important in P1, so we suspect that
the dust that gives rise to the continuum emission is likely
to come mainly from the outer layers of molecular clouds
exposed to the starlight.  The relative importance of
large-grain (equilibrium) heating and small-grain (transient)
heating is difficult to establish solely with the new data
presented here.

\section{Conclusions\label{ch3:c}}

We have presented spatially resolved 8--13 \mum spectroscopy of
the nuclear region of the starburst galaxy NGC 253.  We find
that high-mass star formation explains the observed spectra
through a combination of PAH emission, [\ion{Ne}{2}] emission,
and thermal continuum with little need to invoke absorption by
cold dust greater than A$_V\sim$15.

Based on the revised astrometry of Kalas \&
Wynn-Williams\markcite{n253-ka94} (1994) we identify P1 with
an optical red object observed by Watson et al. (1996) and
with the 2-cm radio source TH7 observed by Turner \& Ho (1995).

We report the measurement of a low-contrast
but high-intensity plateau of PAH emission in the 11.1--12.9
\mum range, and
identify it with emission observed in PDRs in the vicinity
galactic \ion{H}{2} 
regions.  

We set a limit on the effective temperature of the intergrated emission from
stars sampled by our spectra to be less than 36,000 K based on
the comparison of I([\ion{Ne}{2}]) and upper limits on
I([\ion{S}{4}])  under the assumption that the dust optical
depth is not very large at 10.5 \micron.

We compare I([\ion{Ne}{2}]) with I(Br$\gamma$) and radio continuum
emission from P1, and find reasonable agreement with standard
recombination theory, while for P2a the radio continuum emission is
probably too strong to be due to \ion{H}{2} regions.

We examine the relative intensities of 11.3-\mum PAH emission feature,
plateau emission, and the [\ion{Ne}{2}] emission line, and the
continuum intensity and color temperature and find that while
the [\ion{Ne}{2}] and 11.3 \mum PAH emission are roughly
proportional, as observed in other starburst galaxies, the 11.3
\mum PAH feature and plateau emission show a larger range of
variation with respect to each other.  We suggest that 
enhanced plateau emission is due to the predominance of
conditions more similar to the depths of PDRs than regions
adjacent to ionization fronts.

\noindent{ACKNOWLEDGMENTS}

We would like to thank the UKIRT staff--Joel Aycock, Dolores
Walther, and Thor Wold--for assistance at the telescope, and
Tom Geballe, for helpful discussions.  UKIRT is operated by
the Joint Astronomy Centre on behalf of the U.K.~Particle
Physics and Astronomy Research Council.  F. Crifo gave us
very helpful advice concerning the comparative reliability of
different astrometric catalogs.  Many members of the
Institute for Astronomy have benefited this work through
discussion and sharing of work in progress.  A few are Dave
Sanders, Bob Joseph, Klaus Hodapp, Alan Tokunaga, Joe Hora,
Tom Greene, Paul Kalas, and Jason Surace; we thank these and
others.  This work has been partially supported by NSF grant
ASTR-8919563, and NASA grant NAGW-3938.  The NASA
Extragalactic Database, (NED) has been a constant aid.  It is
run by the Jet Propulsion Laboratory, California Institute of
Technology, under contract with the National Aeronautics and
Space Administration.  This work has also made use of NASA's
Astrophysics Data System Abstract Service and Catalog database.

\newpage

{
\begin{deluxetable}{lcccc}
\tablewidth{0pt}
\tablecaption{Observing Log for NGC 253}
\tablecolumns{5}
\tablehead{Date & \colhead{Aperture} & \colhead{Sources} & \colhead{Time} & \colhead{Offset}\\
(UT) & \colhead{(\arcsec~)} & \colhead{Observed} & \colhead{on Source (s)} & \colhead{from P1} }
\startdata
 1992 Oct 15 & 3.26& P1\phantom{a} & 200 &   \nl
 & & P2b & 500 & 2\farcs7E 2\farcs0N\nl
 1993 Aug 31& 2.40& P0\phantom{a} & 500 & 1\farcs8W 2\farcs0S \nl
 & & P1\phantom{a} & 350 & \nl
 & & P2a & 500 & 1\farcs8E 2\farcs0N \nl 
 1994 Aug 28& 3.26& P1\phantom{a} & 100 & \nl
\enddata
\end{deluxetable}

\begin{deluxetable}{lcccc}
\tablewidth{0pt}
\tablecaption{Positions of Bright Peaks in NGC 253}
\tablecolumns{5}
\tablehead{Publication & \colhead{$\lambda$} & \colhead{Source Name} & \multicolumn{2}{c}{
Position (1950)} \\
 &  &  & \colhead{$\alpha$ = 00$^{\rm h}$ 45$^{\rm m}$} & \colhead{$\delta$ = $-$25$^\circ$ 33\arcmin} }
\startdata
Kalas \& Wynn-Williams & 4.8 \micron & Peak 1 & 05\fs632 $\pm$ 0.037 &
41\farcs34 $\pm$ 0.5 \nl
Turner \& Ho & 2 cm & 7 & 05\fs626\phantom{ $\pm$ 0.020} & 41\farcs23\phantom{ $\pm$ 0.2} \\
Watson et al. & 0.6 \mum & Bright blob & 05\fs58\phn $\pm$ 0.02\phn & 41\farcs08 $\pm$ 0.3 \nl
\nl
Kalas \& Wynn-Williams & 4.8 \micron & Peak 2 & 05\fs782 $\pm$ 0.037 &
39\farcs50 $\pm$ 0.5 \nl
Turner \& Ho & 2 cm & 2 & 05\fs795\phantom{ $\pm$ 0.022} & 38\farcs95\phantom{ $\pm$ 0.2} \\
\nl
Kalas \& Wynn-Williams & 4.8 \micron & Peak 3 & 06\fs08\phn $\pm$ 0.037 &
35\farcs89 $\pm$ 0.5 \nl
Watson et al. & 0.6 \mum & Spot $a$ & 06\fs03\phn $\pm$ 0.02\phn & 35\farcs61 $\pm$ 0.3 \nl
\cline{4-5} \nl
{\em Hipparcos} & \nodata & SAO 166579  & $\alpha$ = 00$^{\rm h}$ 45$^{\rm m}$ & $\delta$
 = $-$25$^\circ$ 40\arcmin \nl
& & & 7\fs4638 $\pm$ 0.0004 & 3\farcs677 $\pm$ 0.005 \nl
\enddata
\end{deluxetable}

\begin{deluxetable}{lcccc}
\tablewidth{0pt}
\tablecaption{Spectral Measurements of NGC 253}
\tablecolumns{5}
\tablehead{ & \multicolumn{4}{c}{Position} \\
\cline{2-5}\\
Property & \colhead{P0} & \colhead{P1}& \colhead{P2a}& \colhead{P2b}}
\startdata
 T$_{\rm cont}$ [K]  & 119$^{+9}_{-8}$  & 145$^{+2}_{-2}$& 136$^{+4}_{-4}$   & 119$^{+5}_{-5}$    \nl
 $F_{\nu}$(13.2 $\mu$m) [Jy]  & 1.7 $\pm$ 0.1  & 6.7 $\pm$ 0.1 & 4.3 $\pm$ 0.1  & 3.3 $\pm$ 0.1    \nl
 I$_{11.3\mu {\rm m}}$ [$10^{-15}$ W m$^{-2}$]  & 3.3 $\pm$ 0.4  & 6.0 $\pm$ 0.3 & 6.7 $\pm$ 0.3  & 5.9 $\pm$ 0.4   \nl
 I$_{\rm [NeII]}$ [$10^{-15}$ W m$^{-2}$]  & 3.9 $\pm$ 0.8  & 12.2 $\pm$ 0.8 & 9.0 $\pm$ 0.8 & 10.9 $\pm$ 0.9   \nl
 I$_{11.1-12.9\mu {\rm m}}$ [$10^{-15}$ W m$^{-2}$]  &  6 $\pm$  2  &  30 $\pm$  2  &  25 $\pm$  2 &  17 $\pm$  2    \nl
 I$_{\rm [ArIII]}$ [$10^{-15}$ W m$^{-2}$]  & $<$1.3  & $<$1.1 & $<$1.3  & $<$1.3   \nl
 I$_{\rm [SIV]}$ [$10^{-15}$ W m$^{-2}$]  & $<$1.3  & $<$1.1 & $<$1.3  & $<$1.3  \nl
 EW 11.3 [$\mu$m]  & 0.22 $\pm$ 0.02  & 0.075 $\pm$ 0.004 & 0.14 $\pm$ 0.01 & 0.19 $\pm$ 0.01  \nl
 EW [\ion{Ne}{2}] [$\mu$m]  & 0.13 $\pm$ 0.03  & 0.09 $\pm$ 0.01 & 0.11 $\pm$ 0.01  & 0.17 $\pm$ 0.01  \nl
 ${I_{11.1-12.9\mu {\rm m}} \over \rm I_{11.3\mu m}}$  & 2.0 $\pm$ 0.6  & 5.0 $\pm$ 0.5 & 3.8 $\pm$ 0.3 & 2.9 $\pm$ 0.3  \nl
 $\tau_{\rm Sil}$(9.7)  & $<$1.3  & $<$0.7 & $<$0.8  & $<$1.6    \nl
\enddata
\end{deluxetable}
}

\begin{figure}
\caption[NGC 253 at 5 \micron]{The positions and apertures for
the spectra presented in this paper are shown as open circles and
labeled P0, P1, P2a, and P2b.  The filled circles
are the positions of compact radio sources measured by Turner Ho
(1985), with the larger two labeled according to their table 2.
Kalas \& Wynn-Williams's Peak 3 is also labeled. This map is
adapted from the 4.8 \mum map of Kalas \& Wynn-Williams (1994).}

\end{figure}
\begin{figure}
\caption[The 8--13 \mum Spectra at Four Positions in NGC 235]{The 8--13 \mum spectra at four positions in the
starburst nucleus of NGC 253 are labeled as in
Table 1. Large filled circles represent observed data points,
while small filled circles, when present, indicate boxcar
smoothed data, where the smoothing radius is 0.1 \micron. The
solid lines are blackbody fits between 10 and 11 \mum and at
13 \micron.  The temperature of the fitted blackbody is
indicated in each panel. The dashed lines show a fit of the
Orion Bar spectrum (Roche 1989) between 10.9 and 11.7 \micron,
while the shaded regions indicate the expected range in the
strength of the 7.7 \mum feature based on the strength of the
11.3 \mum feature.  }
\end{figure}

\begin{figure}
\caption[The Continuum Subtracted 10--13.3 \mum 
Spectra]{ The continuum subtracted 10--13.3 \mum 
spectra corresponding to Fig. 2 are presented. 
The filled circles are the continuum-subtracted 
data displayed on linear scales in $F_\lambda$ 
and wavelength, with 1 $\sigma$ errors that do not 
include uncertainties in the continuum subtraction.
The solid lines give the total model fits for each spectrum.
The dot-dashed line in the P0 spectrum indicates 
the contribution of the plateau and 12.7 \mum feature 
that are subtracted from the [\protect\ion{Ne}{2}] flux.}
\end{figure}

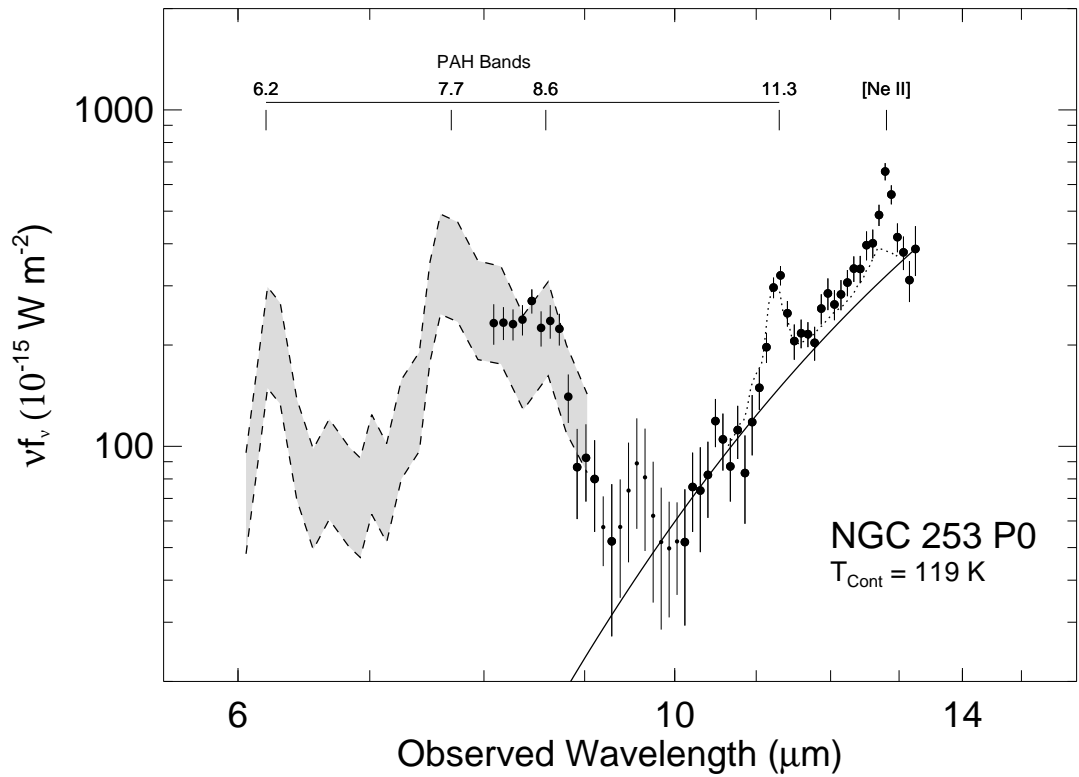
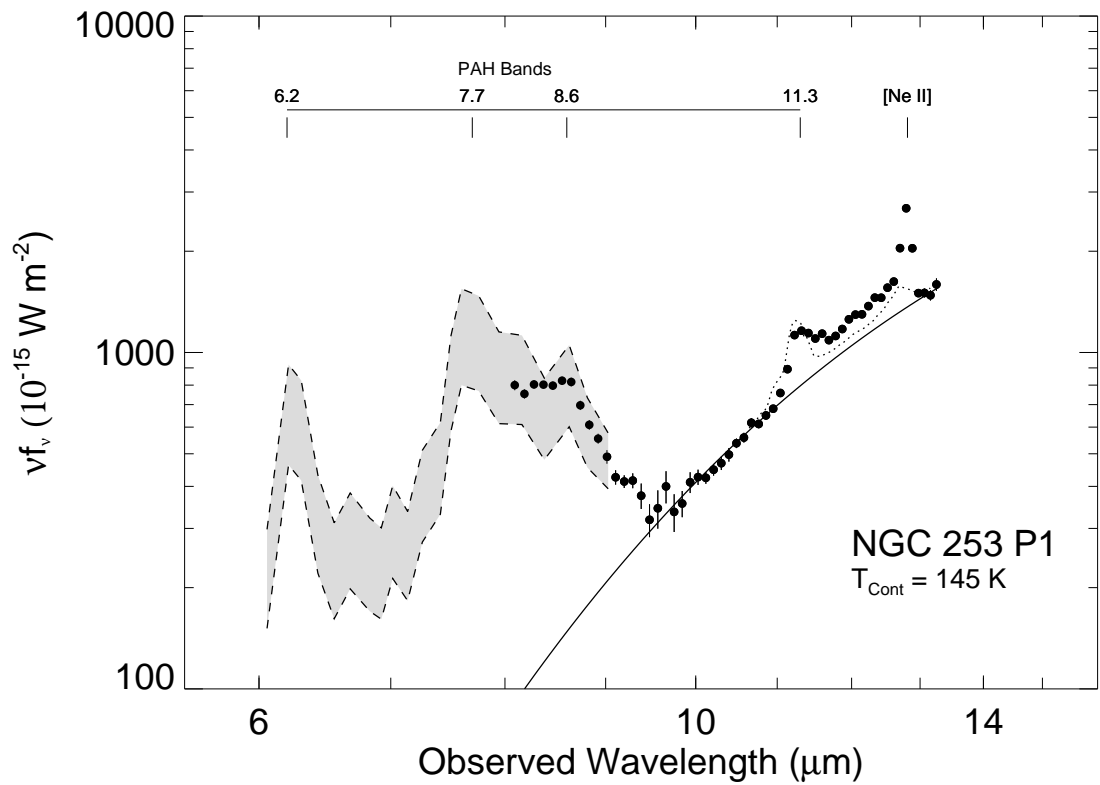
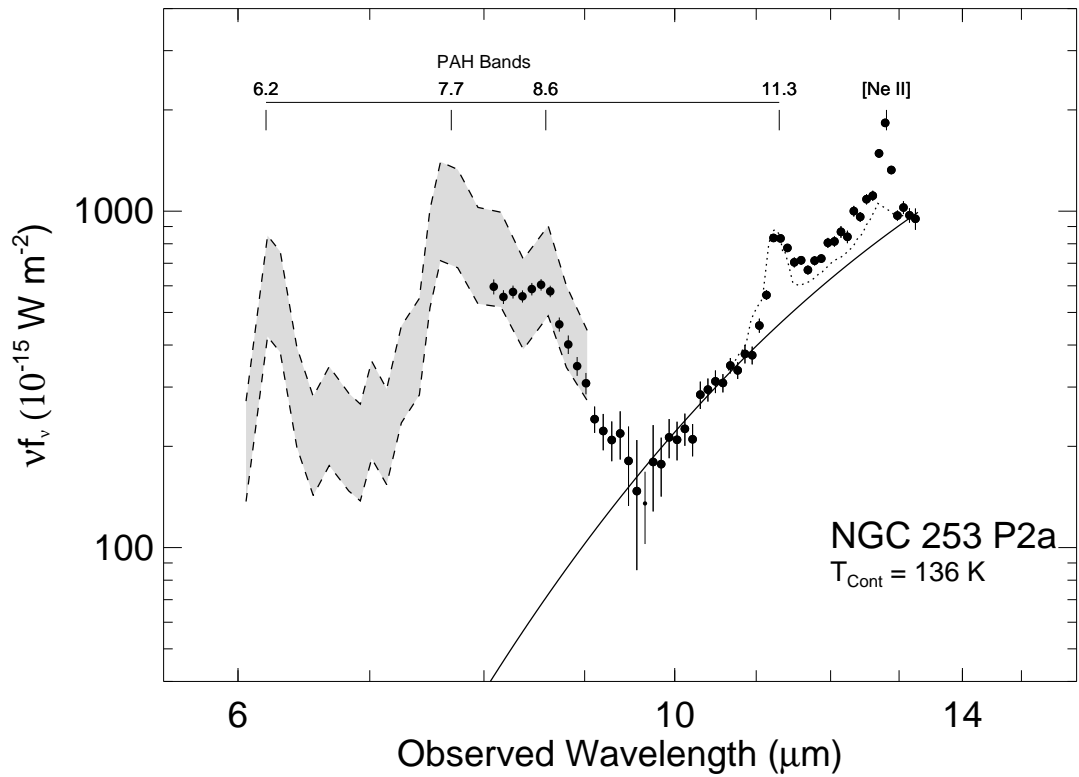
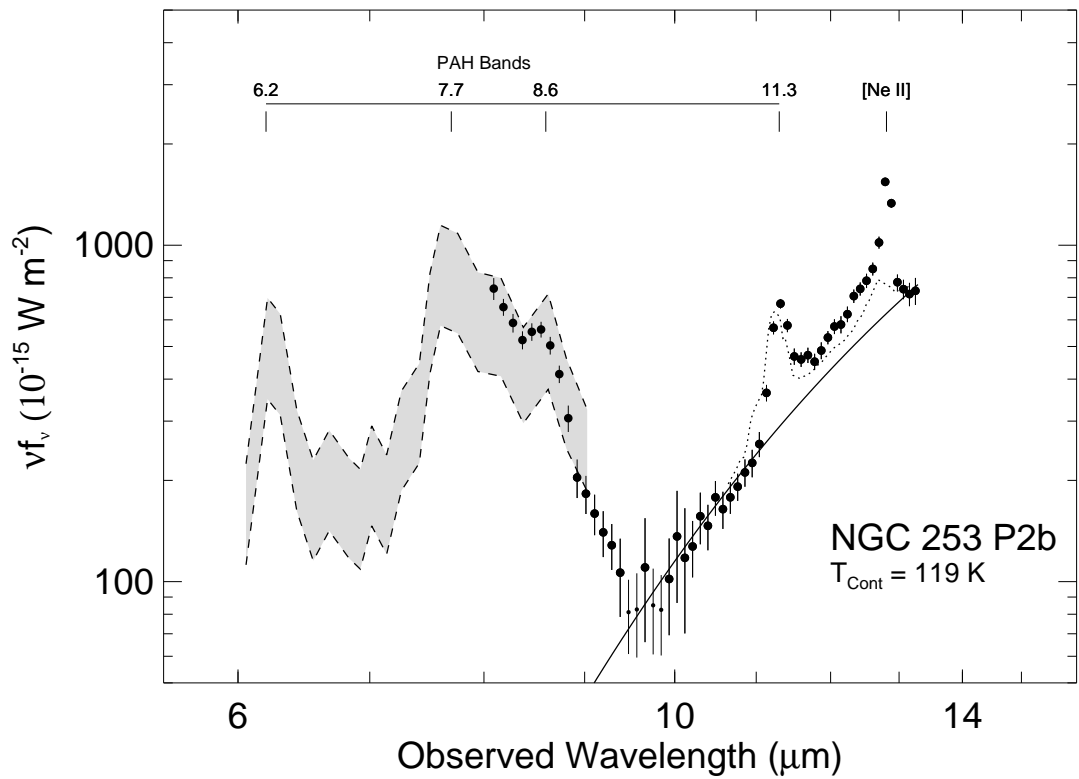
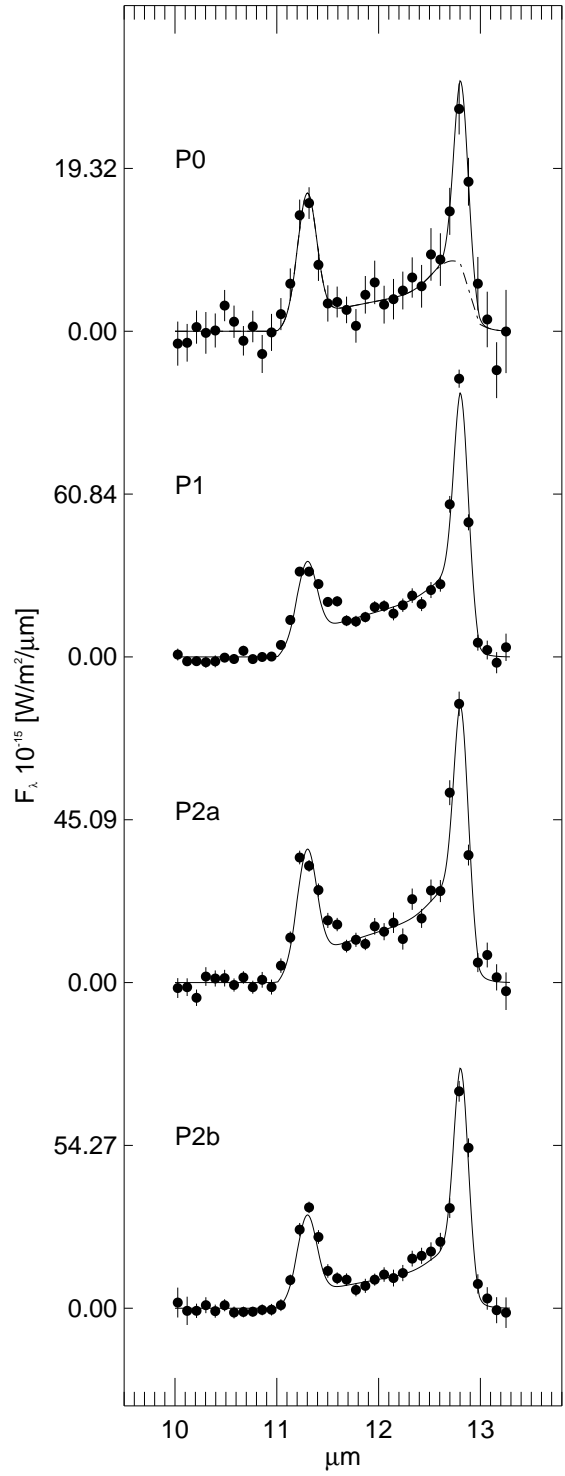
\end{document}